\documentclass[prl,twocolumn,preprintnumbers,superscriptaddress]{revtex4}

\usepackage{epsf}
\usepackage{natbib}
\usepackage{dcolumn}
\usepackage{bm}
\usepackage{ulem}
\usepackage{amsmath}
\usepackage{amssymb}
\usepackage{color}
\usepackage{latexsym}
\usepackage{graphicx}
\usepackage{dcolumn}
\usepackage{bm}
\usepackage[
colorlinks=true,
citecolor=blue,
setpagesize=false]{hyperref}

\begin{document}

\title{Pressure-induced phase switching of Shubnikov de Haas oscillations in the molecular Dirac fermion system $\alpha-$(BETS)$_{2}$I$_{3}$}
\author{Yoshitaka Kawasugi} \email{yoshitaka.kawasugi@sci.toho-u.ac.jp}
\affiliation{Department of Physics, Toho University, Funabashi, Chiba 274-8510, Japan}
\affiliation{RIKEN, Wako, Saitama 351-0198, Japan}
\author{Hikaru Masuda}
\affiliation{Department of Physics, Toho University, Funabashi, Chiba 274-8510, Japan}
\author{Masashi Uebe}
\affiliation{RIKEN, Wako, Saitama 351-0198, Japan}
\author{Hiroshi M. Yamamoto}
\affiliation{RIKEN, Wako, Saitama 351-0198, Japan}
\affiliation{Research Center of Integrative Molecular Systems (CIMoS), Institute for Molecular Science, National Institutes of Natural Sciences, Okazaki, Aichi 444-8585, Japan}
\author{Reizo Kato}
\affiliation{RIKEN, Wako, Saitama 351-0198, Japan}
\author{Yutaka Nishio}
\affiliation{Department of Physics, Toho University, Funabashi, Chiba 274-8510, Japan}
\author{Naoya Tajima} \email{naoya.tajima@sci.toho-u.ac.jp}
\affiliation{Department of Physics, Toho University, Funabashi, Chiba 274-8510, Japan}


\begin{abstract}
We report on the Shubnikov de Haas (SdH) oscillations in the quasi two-dimensional molecular conductor $\alpha-$(BETS)$_{2}$I$_{3}$ [BETS: bis(ethylenedithio)tetraselenafulvalene] laminated on polyimide films at 1.7 K.
From the SdH phase factor, we verified experimentally that the material is in the Dirac fermion phase under pressure.
$\alpha-$(BETS)$_{2}$I$_{3}$ is in the vicinity of the phase transition between strongly correlated insulating and Dirac fermion phases, and is a possible candidate for an ambient-pressure molecular Dirac fermion system.
However, the SdH oscillations indicate that the Berry phase is zero at ambient pressure.
Under pressure, a $\pi$ Berry phase emerges when the metal-insulator crossover is almost suppressed at $\sim$0.5 GPa.
The results contrast those for the pioneering molecular Dirac fermion system $\alpha-$(BEDT-TTF)$_{2}$I$_{3}$ [BEDT-TTF: bis(ethylenedithio)tetrathiafulvalene] in which Dirac fermions and semiconducting behavior are simultaneously observed.
%
\end{abstract}

\maketitle

\section*{Introduction}
Dirac electrons in solids, which obey linear (pseudo-relativistic) dispersion relations, are one of the central issues in condensed matter physics, particularly since the experimental discovery of graphene \cite{Novoselov2004}.
However, materials in which the Fermi energy ($E_{\rm F}$) lies at the contact point are still few.
Among them, the molecular Dirac fermion system $\alpha-$(BEDT-TTF)$_{2}$I$_{3}$ has provided a unique platform for two-dimensional massless Dirac fermions \cite{Tajima2006,Kajita2014}.
Unlike graphene, $\alpha-$(BEDT-TTF)$_{2}$I$_{3}$ is a bulk quasi-two-dimensional material, in which $E_{\rm F}$ is close to the contact points between highly tilted Dirac cones at non-symmetric k-points in the Brillouin zone \cite{Katayama2006}.
The low Fermi velocity ($\sim 10^{4}$ m/s) and the low damping of the Landau levels allow us to precisely investigate how the Landau levels form and separate towards the quantum limit \cite{Kobara2020}.

\begin{figure}[htbp]
  \begin{center}
    \includegraphics{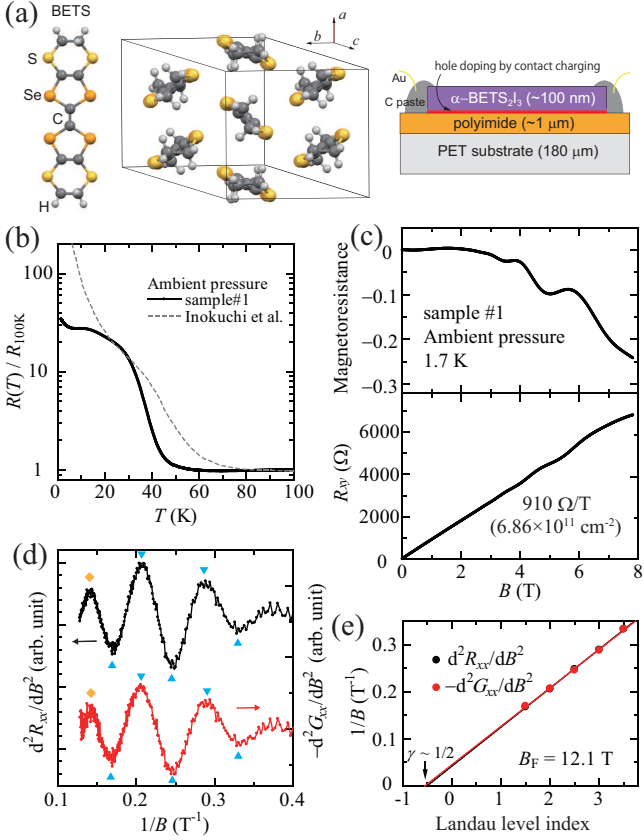}
  \caption{(a) Molecular arrangement of $\alpha-$(BETS)$_{2}$I$_{3}$ (I$_{3}^{-}$ is not shown) and schematic side view of the sample. (b) Temperature dependence of the resistivity at ambient pressure in sample \#1. Dotted line denotes the data of a bulk crystal of $\alpha-$(BETS)$_{2}$I$_{3}$ in the literature \cite{Inokuchi1995}. (c) Magnetic field dependence of the longitudinal magnetoresistance $(R_{xx}(B)-R_{xx}(0))/R_{xx}(0)$ and the Hall resistance $R_{xy}$ at 1.7 K in sample \#1. (d) $1/B$ dependences of $d^{2}R _{xx}/dB^{2}$ ($\propto -\Delta R_{xx}$) and $-d^{2}G_{xx}/dB^{2}$ ($\propto \Delta G_{xx}$) derived from the data in Fig. 1(c). Blue triangles indicate minima and maxima. Yellow diamonds denote peaks suspected to be Zeeman splitting peaks (see text). (e) Landau fan diagram constructed from Fig. 1(d).}
  \end{center}
\end{figure}

\begin{figure*}[htbp]
  \begin{center}
    \includegraphics{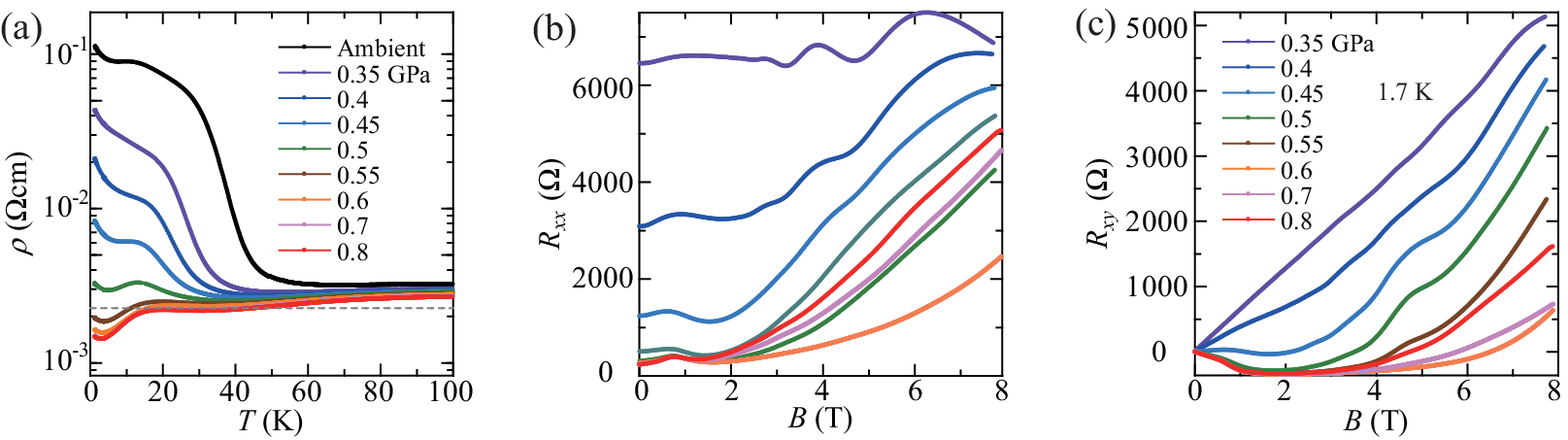}
  \caption{Pressure dependence of the resistivity in sample \#1. (a) Temperature dependence of the resistivities. Dashed line indicates the resistivity when each conducting layer has the quantum resistance $h/e^{2}$. (b),(c) Magnetic field dependences of $R_{xx}$ and $R_{xy}$.
  }
  \end{center}
\end{figure*}

The massless Dirac fermion phase appears in the vicinity of a strongly correlated insulating phase by application of pressure above 1.5 GPa.
Therefore, the interaction effect on the massless Dirac fermions in this system has also been of great interest, and peculiar phenomena, such as an anisotropic Dirac cone reshaping due to the tilt of the cone \cite{Hirata2016} and ferrimagnetic spin polarization due to short-range Coulomb interaction \cite{Hirata2017}, have been reported.
In addition, a deviation from the Korringa law in NMR measurement suggests that the system is in the strong coupling regime that graphene cannot reach \cite{Hirata2017}.
Probably because of these special situations, the insulating behavior and charge gap remain even in the massless Dirac fermion phase under high pressure \cite{Liu2016,Beyer2016,Uykur2019}.
The short-range interaction effect may become even more significant in the vicinity of the correlated insulating phase.
Recently, the quantum phase transition between the insulating phase and the massless Dirac fermion phase was reported \cite{Unozawa2020}.
The Fermi velocity ($v_{\rm F}$) decreases without creating a mass gap upon approaching the phase transition.
Further detailed experiments around the phase transition in $\alpha-$(BEDT-TTF)$_{2}$I$_{3}$ and its related materials will be interesting as there are no other massless Dirac fermion systems in such a strong electron correlation regime.

$\alpha-$(BETS)$_{2}$I$_{3}$ \cite{Inokuchi1993,Inokuchi1995}, the selenium analog of $\alpha-$(BEDT-TTF)$_{2}$I$_{3}$, may be an excellent platform to explore electronic states in the vicinity of the phase transition.
It shows similar resistivity behavior to $\alpha-$(BEDT-TTF)$_{2}$I$_{3}$, but the insulating phase can be suppressed under lower pressure (0.6 GPa), probably due to the large bandwidth \cite{Tajima2006}.
The transport properties above 0.6 GPa are reminiscent of those of $\alpha-$(BEDT-TTF)$_{2}$I$_{3}$ above 1.5 GPa.
Therefore, the electronic state of $\alpha-$(BETS)$_{2}$I$_{3}$ is considered similar to that of $\alpha-$(BEDT-TTF)$_{2}$I$_{3}$ at approximately 0.9 GPa.
Indeed, band calculations based on the crystal structure under high pressure indicate the presence of the Dirac cones in both $\alpha-$(BEDT-TTF)$_{2}$I$_{3}$ and $\alpha-$(BETS)$_{2}$I$_{3}$ \cite{Kondo2009,Alemany2012} (although Dirac and normal electrons coexist).
However, in-depth verification of Dirac fermions with the quantum oscillation measurements has not been reported so far.
Recently, first-principles calculations by multiple independent research groups indicate that $\alpha-$(BETS)$_{2}$I$_{3}$ is a type-I Dirac fermion system even at ambient pressure \cite{Ohki2020,Tsumuraya2020,Kitou2020}.
Those groups simultaneously suggest the possibility of different insulating mechanisms from $\alpha-$(BEDT-TTF)$_{2}$I$_{3}$ (spin-orbit coupling by Kitou et al. \cite{Kitou2020}, and Coulomb interaction + spin-orbit coupling by Ohki et al. \cite{Ohki2020}).

In this study, to uncover the presence of Dirac cones, with similarities to and differences from $\alpha-$(BEDT-TTF)$_{2}$I$_{3}$ in the vicinity of the phase transition, we investigate the Shubnikov de Haas (SdH) oscillation in thin single crystals of $\alpha-$(BETS)$_{2}$I$_{3}$ laminated on polyimide films using a similar experimental method to $\alpha-$(BEDT-TTF)$_{2}$I$_{3}$ \cite{Tajima2013}.
From these measurements, we verified that $\alpha-$(BETS)$_{2}$I$_{3}$ is in the Dirac fermion phase under pressure.
The period of the oscillation does not significantly change before and after the transition, indicating that $\alpha-$(BETS)$_{2}$I$_{3}$ under pressure has no large Fermi surfaces (type-I Dirac fermion system).
Under high pressure, $\alpha-$(BETS)$_{2}$I$_{3}$ is in the Dirac fermion phase with approximately 20\% lower Fermi velocity than that in similarly doped $\alpha-$(BEDT-TTF)$_{2}$I$_{3}$ \cite{Unozawa2020}.

\section*{Methods}
Polyimide films (CT4112, KYOCERA Chemical Corporation) were spin-coated on polyethylene terephthalate (PET) substrate (Teflex FT7, Teijin DuPont Films Japan Limited) and baked at 180 $^{\circ}$C for 1 hour.
We electrochemically synthesized a thin ($\sim$100 nm) single crystal of $\alpha-$(BETS)$_{2}$I$_{3}$ from a chlorobenzene solution (2\% v/v methanol) of BETS \cite{Kato1991} and tetrabutylammonium triiodide by applying 5 $\mu$A for 20 hours.
The thin crystal was transferred into 2-propanol with a pipette and guided onto the substrate.
After the substrate was removed from the 2-propanol and dried, the crystal naturally adhered to the substrate.
The x-ray diffraction measurement is difficult because the crystal is thin and laminated on the noncrystalline polymer substrate.
However, thanks to the polarizing property of I$_{3}^{-}$, optical images through a polarizer indicate that the crystal is a single crystal in which the two-dimensional conducting plane is parallel to the substrate \cite{Supplemental}.
The $a$ and $b$ axes tend to correspond to the diagonals of the crystal if the shape is close to diamond.
Atomic force microscopy revealed that the surface roughness of the crystal was smaller than the thickness of the BETS conducting layer \cite{Supplemental}.

Unlike $\alpha-$(BEDT-TTF)$_{2}$I$_{3}$, no polymorphs of (BETS)$_{2}$I$_{3}$ have been reported, and the temperature dependence of the resistance is similar to the literatures \cite{Inokuchi1993,Inokuchi1995} (as shown later).
We made electrical contacts with carbon paste and Au wires.
Samples \#1 and \#2 were subsequently shaped into Hall bars using a pulsed laser beam with a wavelength of 532 nm (sample \#3 and \#4 were not shaped).
The dimensions of samples \#1-4 are 90 $\mu$m (width)$\times$ 180 $\mu$m (length) $\times$ 130 nm (thickness), 90 $\mu$m $\times$ 110 $\mu$m $\times$ 90 nm, 310 $\mu$m $\times$ 160 $\mu$m $\times$ 80 nm, and 130 $\mu$m $\times$ 130 $\mu$m $\times$ 125 nm, respectively.

For samples \#1-3, we measured the longitudinal resistivity and the Hall resistivity using a dc current of 1 $\mu$A from a dc source (KEITHLEY 2400, Keithley Instruments) and a nano voltmeter (Agilent 34420A, Agilent Technologies) in a cryostat with a superconducting magnet that generate up to 8 T (TeslatronPT, Oxford Instruments).
For sample \#4, a dc current of 10 $\mu$A was applied from a dc source (KEITHLEY 6221, Keithley Instruments) and a nano voltmeter (Agilent 34420A, Agilent Technologies) in a He$^{3}$ cryostat with a superconducting magnet that generate up to 10 T (Cryogenic Limited).
The magnetic field was applied perpendicular to the substrate of the samples.
For pressure measurements, we employed a typical CuBe pressure cell and Daphne 7373 oil.
The pressures are values at room temperature, and the actual pressures at low temperatures are 0.1$\sim$0.2 GPa less than the notations \cite{Murata1997}.

Besides, the polymer film is not restricted to polyimide.
We also observe similar SdH oscillations in $\alpha-$(BETS)$_{2}$I$_{3}$ directly laminated on the PET substrate.
However, we employed polyimide films because the oscillation signals tended to be more clear probably due to more clean surface conditions of our polyimide films.

\section*{Results and Discussion}

The SdH oscillation is a powerful tool to investigate the Fermi surface and the Berry phase \cite{Mikitik1999}.
Neither $\alpha-$(BEDT-TTF)$_{2}$I$_{3}$ nor $\alpha-$(BETS)$_{2}$I$_{3}$ shows the SdH oscillations in their bulk crystals regardless of pressure.
We have to dope some carriers to observe the oscillations.
Here, we synthesize thin single crystals of $\alpha-$(BETS)$_{2}$I$_{3}$ and laminate them on polyimide films.
The contact charging between $\alpha-$(BETS)$_{2}$I$_{3}$ and polyimide induces hole doping, resulting in the observation of the SdH oscillations (one or two conducting layers are doped in the case of $\alpha-$(BEDT-TTF)$_{2}$I$_{3}$ \cite{Tajima2013}).
According to the SdH oscillations period, the hole density is approximately 10$^{12}$ cm$^{-2}$, corresponding to $\sim$0.5\% of the first Brillouin zone.
Notice that the thin crystal consists of several tens of conducting BETS and insulating I$_{3}$ layers (as shown in the Methods section), but the doped carriers are confined at the surface.
Therefore, the sample resistance is the combined resistance of the nondoped bulk and doped surface, and is difficult to separate.
Nevertheless, we can investigate the doped surface using the SdH oscillations because the nondoped bulk does not show the oscillations.
The application of contact charging also causes unintended strain effects from the substrate.
The crystal of the target material is much thinner than the substrate and tightly adheres to the substrate.
Therefore, thermal and mechanical contractions (due to cooling and pressure) of the nondoped bulk ($\sim 100$ nm) and the doped surface (a few nanometers) are governed by those of the substrate.
These effects modify the effective pressure of the laminated crystal, as shown later.
However, this effect does not change the essential pressure effect because the strain is biaxial and parallel to the conducting plane.
If we employ an unshrinkable substrate such as Si, the shrinkable molecular crystal is broken under pressure (probably due to the Poisson effect).
We employed shrinkable plastic substrates in this study.

The period and phase of the oscillations imply the following.
At ambient pressure, the charge carriers are not Dirac fermions at the doping levels in this study.
The charge carriers turn out to be Dirac fermions when the metal-insulator crossover is sufficiently suppressed by applying pressure.
The phase switching contrasts the behavior in $\alpha-$(BEDT-TTF)$_{2}$I$_{3}$ in the intermediate pressure region, which shows a $\pi$ Berry phase along with insulating behavior \cite{Unozawa2020}.

\subsection*{Ambient pressure}
Figure 1(b) shows the temperature dependence of the resistivity at ambient pressure in sample \#1 ($\alpha-$(BETS)$_{2}$I$_{3}$/polyimide/PET).
Compared with a bulk crystal \cite{Inokuchi1995}, the sample exhibits slightly lower metal-insulator crossover temperatures and more moderate resistivity increases at lower temperatures.
The former is ascribable to the fact that the thermal contraction of the PET substrate applies compressive strain to $\alpha-$(BETS)$_{2}$I$_{3}$ \cite{Kawasugi2008} (therefore, this sample at ambient pressure corresponds to a bulk crystal under weak pressure), and the latter is attributable to the doping effect of the polyimide layer.
As the single crystal consists of several tens of conducting (BETS) layers and the doped carriers are confined at the interface, the doping effect appears only at low temperatures where the bulk is insulating.
At 1.7 K, the sheet resistivity (resistance $\times$ width $/$ length) is 8.4 $\times$ 10$^{3}$ $\Omega$.

Figure 1(c) shows the magnetoresistance (upper) and the Hall resistance (lower) at 1.7 K.
The SdH oscillations along with negative magnetoresistance are visible.
The magnetoresistance is complicated in detail.
It is slightly negative up to 0.4 T, turns positive up to 1.5 T, and then becomes negative again by a further magnetic field.
Such a negative magnetoresistance has not been observed in $\alpha-$(BEDT-TTF)$_{2}$I$_{3}$ under low pressures \cite{Unozawa2020}.
The magnetoresistance can be simply explained by neither the weak localization nor weak-antilocalization.
It is reminiscent of the negative longitudinal magnetoresistance in topological semimetals \cite{Schumann2017} due to charge carrier density or mobility fluctuations.
However, the magnetic field direction is different in this study (current is perpendicular to the magnetic field).
Its origin cannot be clarified at this moment.
We observe the negative magnetoresistance (without oscillations) in a bulk crystal \cite{Supplemental}.
Although we cannot see whether the doped interface also shows the negative magnetoresistance or not, the oscillations originate from the interface.
We focus on the oscillation signals in this study.

The low-field Hall resistances are positive and proportional to the magnetic field.
The sign becomes negative at around 23 K with increasing temperature \cite{Supplemental}.
By contrast, a bulk crystal shows negative Hall resistance at low temperatures \cite{Supplemental}.
Therefore, the doped carriers are holes and the concentration is $\sim$10$^{12}$ cm$^{-2}$ by ignoring electrons in bulk (Hall mobility $\sim 1080$ cm$^{2}$/Vs).

The quantum oscillations originate from the quantization condition for the energy levels of the electron \cite{Onsager1952,Lifshitz1956}:
\begin{equation}
S_{n}=\frac{2 \pi e}{\hbar}B(n+\gamma),
\end{equation}
where $S_{n}$ is the area of the cyclotron orbit in k-space, $n$ is an integer, $B$ is the magnetic field, and $\gamma$ is the phase factor.
The oscillation signal is periodic against 1/$B$, and the area can be estimated using the measurement $B_{F}\equiv (\frac{1}{B_{n+1}}-\frac{1}{B_{n}})^{-1}$.
Assuming that the spin and valley degeneracies are both 2, the carrier density $N$ is
\begin{equation}
N=\frac{4e}{h}B_{\rm F}.
\end{equation}
The SdH oscillation is given by
\begin{equation}
\Delta R _{xx} = R(B,T) \cos 2\pi (\frac{B_{\rm F}}{B}-\gamma),
\end{equation}
where $R _{xx}$ and $R(B,T)$ are the longitudinal resistance and the oscillation amplitude, respectively \cite{Sharapov2004,Lukyanchuk2004,Zhang2005}.
The phase factor $\gamma$ is associated with the Berry phase $\phi _{\rm B}$ as
\begin{equation}
\gamma-\frac{1}{2}=-\frac{\phi _{\rm B}}{2\pi}.
\end{equation}
In a conventional electron system with isolated bands, $\phi _{\rm B}=0$ and $\gamma=1/2$.
However, if the cyclotron orbit surrounds the contact point of the bands and the energy dispersions are linear in ${\bf k}$ in the vicinity of the contact point, the $\pi$ Berry phase emerges and $\gamma$ becomes zero \cite{Mikitik1999}.
Accordingly, when we plot $1/B$ corresponding to the peaks against Landau level index (Landau fan diagram), the intercept $-\gamma=$ 0 or $-$1/2 for 2D Dirac or normal electrons.

Nevertheless, we have to be careful about the phase analysis of the SdH oscillations.
Eq. (3) assumes the condition $R_{xx}<<|R_{xy}|$ (graphene, $\alpha-$(BEDT-TTF)$_{2}$I$_{3}$, and many low-carrier-density semiconductors meet this condition), and the minima in $R _{xx}$ coincide with those in the conductance $G_{xx}=R_{xx}/(R_{xx}^{2}+R_{xy}^{2})$.
The condition may be violated due to low mobility or the presence of a highly conducting bulk transport channel.
In the case that $R_{xx}>>|R_{xy}|$, the minima in $R_{xx}$ correspond to the maxima in $G _{xx}$.
The difficulty of the phase analysis using resistance data has been pointed out for topological insulators \cite{Xiong2012,Ando2013}.
However, the Hall response is usually weak and the resistance oscillation is more apparent in many cases.
One may still use the reversed resistance data when $R_{xx}>>|R_{xy}|$, as in the case of graphite \cite{Lukyanchuk2004}.
Here, we analyze the oscillations of both $R_{xx}$ and $G_{xx}$ in $\alpha-$(BETS)$_{2}$I$_{3}$/polyimide/PET at ambient pressure.
To eliminate the background, we show the second derivative of the data with respect to the magnetic field.

Figure 1(d) shows the $1/B$ dependences of $d^{2}R _{xx}/dB^{2}$ and $-d^{2}G_{xx}/dB^{2}$ derived from Fig. 1(c).
They correspond to $-\Delta R _{xx}$ and $\Delta G_{xx}$, respectively.
The oscillation signal is more apparent in the upper curve because $\Delta R _{xx}/R>\Delta G_{xx}/\sigma$.
However, both curves show almost the same periods and phases, indicating $-\Delta R_{xx} \propto \Delta G_{xx}$.
$B_{\rm F}$ and $N$ estimated from the upper curve are 12.1 T and $1.17 \times 10^{12}$ cm$^{-2}$, respectively, and correspond to 0.6\% hole doping provided that the doped carriers are confined within one conducting layer.
The $N$ value provides a realistic Hall scattering factor $\gamma _{\rm H}$ of 1.70 ($N=\gamma _{\rm H}/eR_{\rm H}$).
$\gamma$ is almost $1/2$, indicating that the carriers are not Dirac fermions at this doping level.
The same analysis of sample \#2 is described in Fig. S4 \cite{Supplemental}.
The transport properties of sample \#1, such as the temperature dependence of the resistance, the magnetoresistance, the sign and magnitude of the Hall effect, the relationship between resistance and conductance oscillations, and the phase factor reproduced in sample \#2.

The leftmost peak in Fig. 1(d) (denoted by yellow diamonds) deviates from the position predicted from the fitting line in Fig. 1(e).
Provided that this is a split peak as a result of the Zeeman effect, we estimate the effective mass $m^{*} \sim 0.43m_{e}$ from the relation $(n_{\rm LL}+\frac{1}{2})\frac{\hbar e}{m^{*}}B_{\rm LL}=(n_{\rm LL}+\frac{1}{2})\frac{\hbar e}{m^{*}}B_{\rm Z}+\mu _{\rm B}B_{\rm Z}$, where $B_{\rm LL}$ and $B_{\rm Z}$ are the predicted and observed peaks, respectively.

\subsection*{Under pressure}
With increasing pressure, the entire sample is compressed.
The bandwidths of the bulk and surface are enhanced, and their resistances decrease.
The metal-insulator crossover gradually diminishes and disappears at around 0.6 GPa, as shown in Fig. 2(a).
The dip at around 35 K and the upturn below 5 K of the resistance have also been observed in bulk crystals \cite{Inokuchi1993}, but the detailed mechanisms are still unclear.
Above 0.6 GPa, the resistivity is almost constant down to approximately 15 K, below which the metallic behavior of the doped holes appears.
The sheet resistivity per conducting layer is close to the quantum resistance $h/e^{2}$, as in the case of $\alpha-$(BEDT-TTF)$_{2}$I$_{3}$.
The negative magnetoresistance observed at ambient pressure diminishes and becomes positive (Fig. 2(b)).
The Hall resistance also decreases and becomes nonlinear, probably due to the emergence of a conducting bulk transport channel (Fig. 2(c)).
As stated above, the bulk crystal of $\alpha-$(BETS)$_{2}$I$_{3}$ does not show the SdH oscillations even under pressure.
We investigate the doped surface by the analysis of the oscillations.

Figure 3(a) shows the pressure dependence of $-d^{2}G_{xx}/dB^{2}$.
At 0.35 and 0.4 GPa, the minima give $\gamma \sim 1/2$; this tendency is also observed at ambient pressure.
$m^{*}$ values are estimated to be $\sim 0.37m_{e}$ and $0.35m_{e}$, respectively, showing a decreasing trend with pressure.
At 0.45 GPa, we cannot construct a convincing fan diagram because of ambiguous oscillation signals and large background signals.
However, we can see a half-period oscillation (up to $1/B\sim 0.4$ T$^{-1}$), probably indicating the coexistence of anti-phase oscillations.
One possible scenario is the phase separation between the regions with $\gamma$=1/2 and 0.
Above 0.5 GPa, $\gamma$ becomes almost zero, implying the emergence of Dirac fermions, although the oscillations at low $1/B$ are not clear.
The Landau fan diagrams and the pressure dependence of $B_{\rm F}$ and $\gamma$ are summarized in Fig. 3(b) and (c).
The conductance oscillations and Landau fan diagrams of sample \#3 at ambient pressure and 1.2 GPa are also shown in Fig. S5 \cite{Supplemental}.

We cannot observe Dirac fermions unless the metal-insulator crossover is sufficiently suppressed by pressure.
These results are in contrast to those for $\alpha-$(BEDT-TTF)$_{2}$I$_{3}$, in which the Dirac fermions and the semiconducting behavior are simultaneously observed.
Besides, we cannot confirm the coexistence of Dirac and normal electrons (which has been reported for $\alpha-$(BEDT-TTF)$_{2}$I$_{3}$ \cite{Monteverde2013}) in this study.
The SdH oscillations survive beyond the pressure-induced transition and $B_{\rm F}$ does not significantly vary with pressure.
If large Fermi surfaces emerge by applying pressure, as predicted by band calculations in early reports \cite{Kondo2009,Alemany2012}, the doping effect (and accordingly the quantum oscillations) should be obscured by the dense carriers.

The most straightforward interpretation of the phase switching is that the pressure-induced resistive transition is a semiconductor-Dirac fermion system transition. 
Another possible scenario is the pressure-induced merging of the Dirac cones. 
In that case, the number and area of the Fermi surface generally change at the transition. However, Fig. 3 shows that the $B_{\rm F}$ does not significantly change during the transition. 
To consider the merging of the Dirac cone as the origin of the transition, we need a model and conditions consistent with these measurements.

The SdH oscillations become more evident as pressure increases.
Figure 4 shows the magnetotransport properties of sample \#4 at 1.8 GPa and 0.5 K.
Here, the minima of $R_{xx}$ coincide with those of $G_{xx}$ \cite{Supplemental}.
The minima indicate $\gamma=0$ and the interval of the Zeeman splitting peaks (yellow diamonds in Fig. 4(b)) gives effective Fermi velocity $v_{\rm F}\sim$3.6 $\times$ 10$^{4}$ m/s, which is approximately 20\% lower than that from the same analysis for $\alpha-$(BEDT-TTF)$_{2}$I$_{3}$ \cite{Unozawa2020} ($v_{\rm F}$ is estimated from the relation $\sqrt{2e\hbar v_{\rm F}^{2}|n|B_{\rm Zh}}-\mu _{\rm B}B_{\rm Zh}=\sqrt{2e\hbar v_{\rm F}^{2}|n|B_{\rm Zl}}+\mu _{\rm B}B_{\rm Zl}$, where $B_{\rm Zh}$ and $B_{\rm Zl}$ are the peak fields).
The low $v_{\rm F}$ indicates that the Dirac cone is more tilted or more blunted than $\alpha-$(BEDT-TTF)$_{2}$I$_{3}$.
However, we cannot determine the central origin because the estimated $v_{\rm F}$ is average over the orbit in the reciprocal space \cite{Tajima2007}.
The peak around $1/B\sim 0.15$ further separates into two peaks (green squares in Fig. 4(b)).
We assign the bottom between these peaks to the Zeeman splitting peak because a similar $v_{\rm F}$ of 3.6$\times$ 10$^{4}$ m/s is estimated from the bottoms denoted by red circles.
Therefore, the small separation is considered a valley splitting.
We roughly estimate the valley-splitting energy $\Delta _{\rm v}/k_{\rm B}\sim 0.92B$ K using a similar analysis to $v_{\rm F}$, assuming that $\Delta _{\rm v}$ is proportional to the magnetic field.
A relative permittivity $\epsilon$ of $\sim350$ is derived from the relation $\Delta _{\rm v}=\frac{e^3}{\epsilon \epsilon _{0} K \hbar}B$, where $K$ is the distance between the Dirac cones in k-space (approximated by the inverse lattice constant).
In bulk $\alpha-$(BEDT-TTF)$_{2}$I$_{3}$, $\epsilon$ near the Dirac point is estimated to be $\sim190$ from the interlayer magnetoresistance \cite{Tajima2010}.
As the permittivity decreases near the Dirac point, a comparable value is expected at the Dirac point in $\alpha-$(BETS)$_{2}$I$_{3}$.

\begin{figure}[htbp]
  \begin{center}
    \includegraphics{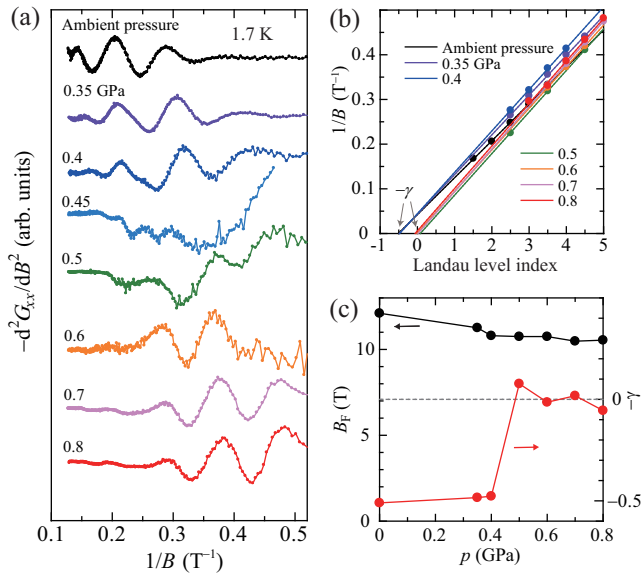}
  \caption{Pressure evolutions of (a) $-d^{2}G_{xx}/dB^{2}$ vs. $1/B$ plots at 1.7 K, (b) Landau fan diagrams, and (c) $B_{\rm F}$ and $\gamma$ in sample \#1.}
  \end{center}
\end{figure}

\begin{figure}[htbp]
  \begin{center}
    \includegraphics{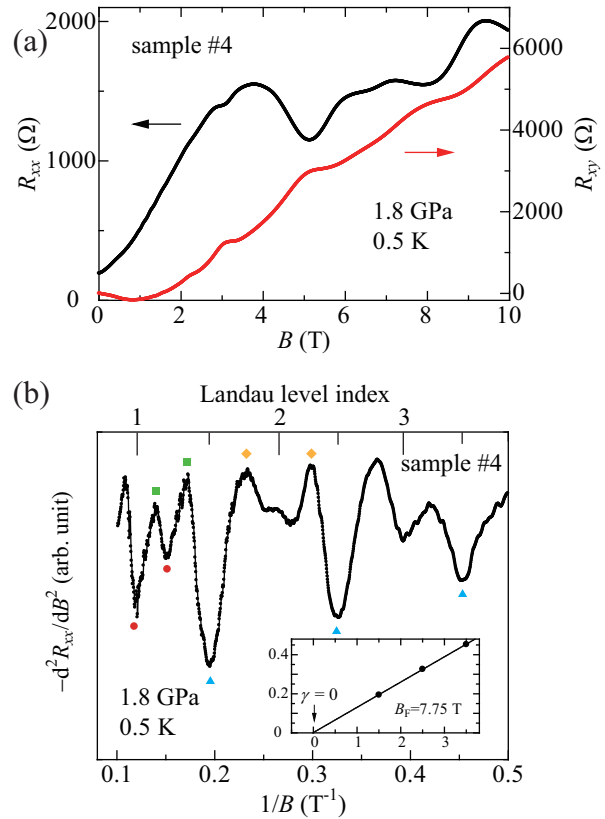}
  \caption{(a) Magnetic field dependences of $R_{xx}$ and $R_{xy}$ in sample \#4 at 1.8 GPa and 0.5 K. (b) $-d^{2}R_{xx}/dB^{2}$ vs. $1/B$ plots derived from Fig. 4(a). Blue triangles, yellow diamonds, green squares, and red circles indicate the minima, the Zeeman splitting peaks at $|n|=2$, the valley splitting peaks, and the bottoms corresponding to the $|n|=1$ Landau level and its Zeeman splitting peak, respectively. Inset shows the Landau fan diagram constructed from the minima in Fig. 4(b). The horizontal and vertical axes are Landau level index and $1/B$, respectively.
  }
  \end{center}
\end{figure}

\subsection*{Summary}
We have investigated the pressure dependence of the magnetoresistance and the Hall effect in slightly hole-doped thin single crystals of $\alpha-$(BETS)$_{2}$I$_{3}$ laminated on polyimide films, and verified that the material is in the Dirac fermion phase under pressure.
We found a phase switching of the SdH oscillation near the pressure-induced metal-insulator crossover, unlike in $\alpha-$(BEDT-TTF)$_{2}$I$_{3}$ in the vicinity of the phase transition.
At ambient pressure, the system exhibits a metal-insulator crossover below 50 K, and the phase of the SdH oscillation at 1.7 K indicates $\gamma=1/2$.
Under pressure, $\gamma$ becomes zero above 0.5 GPa, whereas the metal-insulator crossover disappears at approximately 0.6 GPa.
A half-period oscillation appears at the boundary ($\sim 0.45$ GPa), although the oscillation signal is ambiguous.
It may originate from the coexistence of the regions with normal and Dirac fermions.
The pressure-induced phase switching of the SdH oscillation indicates the presence of a semiconducting phase with normal electrons next to a Dirac fermion phase in $\alpha-$(BETS)$_{2}$I$_{3}$.
In $\alpha-$(BEDT-TTF)$_{2}$I$_{3}$, the $\pi$ Berry phase appears even in the highly resistive states, and such a trivial insulating phase has not been observed \cite{Unozawa2020}.
Recently, Kitou et al. reported that $\alpha-$(BETS)$_{2}$I$_{3}$ maintains the inversion symmetry below the metal-insulator crossover \cite{Kitou2020}, implying a different insulating mechanism from $\alpha-$(BEDT-TTF)$_{2}$I$_{3}$.
Ohki et al. suggested that the insulating phase is a spin-ordered massive Dirac electron phase where time-reversal symmetry is broken but spatial inversion and translational symmetries are conserved \cite{Ohki2020}.
Tsumuraya et al. explained that the system is in the massless Dirac state but a gap opens as a result of the spin-orbit interaction \cite{Kitou2020,Tsumuraya2020}.
However, we cannot confirm the presence of Dirac fermions at ambient pressure in this study.
Under high pressure (1.8 GPa), $\alpha-$(BETS)$_{2}$I$_{3}$ is a Dirac fermion system with $v_{\rm F}$ of $\sim$3.6 $\times$ 10$^{4}$ m/s.
At high magnetic fields, the valley splitting is observed in the SdH oscillation.
The valley splitting energy $\Delta _{\rm v}/k_{\rm B}$ is estimated to be $\sim 0.92B$ K.
Further study is required to clarify the electronic states of $\alpha-$(BETS)$_{2}$I$_{3}$ which may provide a unique Dirac fermion system different from $\alpha-$(BEDT-TTF)$_{2}$I$_{3}$.

\section*{Acknowledgments}
We would like to acknowledge Teijin DuPont Films Japan Limited for providing the PET films. This work was supported by MEXT and JSPS KAKENHI (Grant Nos. JP16H06346, JP19K03730, and JP19H00891).


\begin{thebibliography}{99}
\bibitem{Novoselov2004} K. S. Novoselov, A. K. Geim, S. V. Morozov, D. Jiang, Y. Zhang, S. V. Dubonos, I. V. Grigorieva, and A. A. Firsov, Science {\bf 306}, 666 (2004).
\bibitem{Tajima2006} N. Tajima, S. Sugawara, M. Tamura, Y. Nishio, and K. Kajita, J. Phys. Soc. Jpn. {\bf 75}, 051010 (2006).
\bibitem{Kajita2014} K. Kajita, Y. Nishio, N. Tajima, Y. Suzumura, and A. Kobayashi, J. Phys. Soc. Jpn. {\bf 83}, 072002 (2014).
\bibitem{Katayama2006} S. Katayama, A. Kobayashi, and Y. Suzumura, J. Phys. Soc. Jpn. {\bf 75}, 054705 (2006).
\bibitem{Kobara2020} R. Kobara, S. Igarashi, Y. Kawasugi, R. Doi, T. Naito, M. Tamura, R. Kato, Y. Nishio, K. Kajita, and N. Tajima, J. Phys. Soc. Jpn. {\bf 89}, 113703 (2020).
\bibitem{Hirata2016} M. Hirata, K. Ishikawa, K. Miyagawa, M. Tamura, C. Berthier, D. Basko, A. Kobayashi, G. Matsuno, and K. Kanoda, Nat. Commun. {\bf 7}, 12666 (2016).
\bibitem{Hirata2017} M. Hirata, K. Ishikawa, G. Matsuno, A. Kobayashi, K. Miyagawa, M. Tamura, C. Berthier, and K. Kanoda, Science {\bf 358}, 1403 (2017).
\bibitem{Liu2016} D. Liu, K. Ishikawa, R. Takehara, K. Miyagawa, M. Tamura, and K. Kanoda, Phys. Rev. Lett. {\bf 116}, 226401 (2016).
\bibitem{Beyer2016} R. Beyer, A. Dengl, T. Peterseim, S. Wackerow, T. Ivek, A. V. Pronin, D. Schweitzer, and M. Dressel, Phys. Rev. B {\bf 93}, 195116 (2016).
\bibitem{Uykur2019} E. Uykur, W. Li, C. A. Kuntscher, and M, Dressel, npj Quantum Materials {\bf 4}, 19 (2019).
\bibitem{Unozawa2020} Y. Unozawa, Y. Kawasugi, M. Suda, H. M. Yamamoto, R. Kato, Y. Nishio, K. Kajita, T. Morinari, and N. Tajima, J. Phys. Soc. Jpn. {\bf 89}, 123702 (2020).
\bibitem{Inokuchi1993} M. Inokuchi, H. Tajima, A. Kobayashi, and H. Kuroda, Synth. Met. {\bf 56}, 2495 (1993).
\bibitem{Inokuchi1995} M. Inokuchi, H. Tajima, A. Kobayashi, T. Ohta, H. Kuroda, R. Kato, T. Naito, and H. Kobayashi, Bull. Chem. Soc. Jpn. {\bf 68}, 547 (1995).
\bibitem{Kondo2009} R. Kondo, S. Kagoshima, N. Tajima, and R. Kato, J. Phys. Soc. Jpn. {\bf 78}, 114714 (2009).
\bibitem{Alemany2012} P. Alemany, J.-P. Pouget, and E. Canadell, Phys. Rev. B {\bf 85}, 195118 (2012).
\bibitem{Ohki2020} D. Ohki, K. Yoshimi, and A. Kobayashi Phys. Rev. B {\bf 102}, 235116 (2020).
\bibitem{Tsumuraya2020} T. Tsumuraya and Y. Suzumura, Eur. Phys. J. B {\bf 94}, 17 (2021).
\bibitem{Kitou2020} S. Kitou, T. Tsumuraya, H. Sawahata, F. Ishii, K. Hiraki, T. Nakamura, N. Katayama, H. Sawa, Phys. Rev. B {\bf 103}, 035135 (2021).
\bibitem{Tajima2013} N. Tajima, T. Yamauchi, T. Yamaguchi, M. Suda, Y. Kawasugi, H. M. Yamamoto, R. Kato, Y. Nishio, and K. Kajita, Phys. Rev. B {\bf 88}, 075315 (2013).
\bibitem{Kato1991} R. Kato and H. Kobayashi, Synth. Met. {\bf 42}, 2093 (1991).
\bibitem{Supplemental} See Supplemental Material at [URL will be inserted by publisher] for details of the samples, magnetoresistance in a bulk crystal without substrate, magneto-transport properties in samples \#2 and \#3, and comparison of conductance and resistance oscillations under high pressure in sample \#4.]
\bibitem{Murata1997} K. Murata, H. Yoshino, H. O. Yadav, Y. Honda, and N. Shirakawa, Rev. Sci. Instrum. {\bf 68}, 2490 (1997).
\bibitem{Kawasugi2008} Y. Kawasugi, H. M. Yamamoto, M. Hosoda, N. Tajima, T. Fukunaga, K. Tsukagoshi, and R. Kato, Appl. Phys. Lett. {\bf 92}, 243508 (2008).
\bibitem{Schumann2017} T. Schumann, M. Goyal, D. A. Kealhofer, and S. Stemmer, {\bf 95}, 241113(R) (2017).
\bibitem{Onsager1952} L. Onsager, Philos. Mag. {\bf 43}, 1006 (1952).
\bibitem{Lifshitz1956} I. M. Lifshitz and A. M. Kosevich, Sov. Phys. JETP {\bf 2}, 636 (1956).
\bibitem{Sharapov2004} S. G. Sharapov, V. P. Gusynin, and H. Beck, Phys. Rev. B {\bf 69}, 075104 (2004).
\bibitem{Lukyanchuk2004} I. A. Luk'yanchuk and Y. Kopelevich, Phys. Rev. Lett. {\bf 93}, 166402 (2004).
\bibitem{Zhang2005} Y. Zhang, Y.-W. Tan, H. L. Stormer, and P. Kim, Nature {\bf 438}, 201 (2005).
\bibitem{Mikitik1999} G. P. Mikitik and Yu. V. Sharlai, Phys. Rev. Lett. {\bf 82}, 2147 (1999).
\bibitem{Xiong2012} J. Xiong, Y. Luo, Y. Khoo, S. Jia, R. J. Cava, and N. P. Ong, Phys. Rev. B {\bf 86}, 045314 (2012).
\bibitem{Ando2013} Y. Ando, J. Phys. Soc. Jpn. {\bf 82}, 102001 (2013).
\bibitem{Monteverde2013} M. Monteverde, M. O. Goerbig, P. Auban-Senzier, F. Navarin, H. Henck, C. R. Pasquier, C. M\'{e}zi\`{e}re, and P. Batail, Phys. Rev. B {\bf 87}, 245110 (2013).
\bibitem{Montambauxa2009} G. Montambauxa, F. Pi\'{e}chon, J.-N. Fuchs, and M.O. Goerbig, Euro. Phys. J. B {\bf 72}, 509 (2009).
\bibitem{Morinari2014} T. Morinari and Y. Suzumura, J. Phys. Soc. Jpn. {\bf 83}, 094701 (2014).
\bibitem{Tajima2007} N. Tajima, S. Sugawara, M. Tamura, R. Kato, Y. Nishio, and K. Kajita, Europhys. Lett. {\bf 80}, 47002 (2007).
\bibitem{Tajima2010} N. Tajima, M. Sato, S. Sugawara, R. Kato, Y. Nishio, and K. Kajita, Phys. Rev. B {\bf 82}, 121420(R) (2010).
\end{thebibliography}
\end{document}